\documentclass[conference]{IEEEtran}
\usepackage[utf8]{inputenc}

\usepackage[
    pdftex,
    colorlinks=true,
    linkcolor=black,
    anchorcolor=black,
    citecolor=black,
    filecolor=black,
    menucolor=black,
    runcolor=black,
    urlcolor=black,
    hyperfootnotes=true,
]{hyperref}
\usepackage{overpic}
\usepackage{colortbl}
\usepackage{tango}
\usepackage[super]{nth}
\usepackage{wrapfig}
\usepackage{epigraph}
\usepackage{fontawesome5}
\usepackage{makecell}
\usepackage{pdfpages}
\usepackage{eurosym}
\usepackage{multirow}
\usepackage{enumitem}
\usepackage{url}
\usepackage{acronym}
\usepackage{graphicx}
\usepackage[capitalise]{cleveref}
\usepackage{flushend}
\usepackage[%
    binary-units,
    load-configurations=binary,
    range-units=single,
    per-mode=symbol,
    detect-all,
    forbid-literal-units,
    exponent-product=\cdot,
]{siunitx}[=2021-04-09]
\ifCLASSOPTIONcompsoc
    \usepackage[caption=false,font=normalsize,labelfont=sf,textfont=sf]{subfig}
\else
    \usepackage[caption=false,font=footnotesize]{subfig}
\fi
\usepackage{booktabs}
\usepackage{pdfprivacy}

\usepackage{pgfplots}

\usetikzlibrary{animations}
\usetikzlibrary{arrows.meta}
\usetikzlibrary{arrows}
\usetikzlibrary{backgrounds}
\usetikzlibrary{calc}
\usetikzlibrary{decorations.markings}
\usetikzlibrary{decorations.pathreplacing}
\usetikzlibrary{external}
\usetikzlibrary{fadings}
\usetikzlibrary{fit}
\usetikzlibrary{matrix}
\usetikzlibrary{patterns}
\usetikzlibrary{positioning}
\usetikzlibrary{shadows}
\usetikzlibrary{shapes.geometric}
\usetikzlibrary{shapes.misc}
\usetikzlibrary{shapes.multipart}
\usetikzlibrary{shapes}
\newcommand{\vertCurledArrow}[4]{%
    \draw[#1] (#3.east) .. controls ([xshift=#2]#3.east) and ([xshift=-#2]#4.west) .. (#4.west);
}

\newcommand{\satellite}[3]{%
    \begin{scope}[shift={#1}, rotate=#2, scale=#3]
        \draw[sat] (-4mm,-4mm) rectangle (4mm,4mm);
        \draw[sat] (-2mm,5mm) rectangle (2mm,15mm);
        \draw[sat] (-2mm,-15mm) rectangle (2mm,-5mm);
        \filldraw[sat] (-6mm,4mm) -- (-6mm,-4mm) arc(-45:45:5.657mm) (-6mm,4mm);
    \end{scope}
}

\pgfplotsset{compat=1.17}

\graphicspath{{img/}}

\makeatletter
\newcommand{\IfPackageLoaded}[3]{\ltx@ifpackageloaded{#1}{#2}{#3}}
\makeatother

\AtBeginDocument{%
    \IfPackageLoaded{natbib}{%
        \setlength{\epigraphwidth}{.65\textwidth}
    }{}
}

\AtBeginDocument{%
    \IfPackageLoaded{enumitem}{%
        
    }{}
}

\AtBeginDocument{%
    \IfPackageLoaded{makecell}{%
        
    }{}
}

\newcommand{\mbps}{\mega\bit\per\second}

\newcommand{\gb}{\giga\byte}

\newcommand{\mib}{\mebi\byte}
\newcommand{\gib}{\gibi\byte}

\def\lr/{\textsc{LongRTT}}
\def\g/{\textsc{Goodput}}
\def\terr/{\textsc{Terrestrial}}
\def\sat/{\textsc{Sat}}
\def\satl/{\textsc{SatLoss}}
\def\eut/{\textsc{Eutelsat}}
\def\astra/{\textsc{Astra}}
\def\crosstraffic/{\textsc{CrossTraffic}}

\def\pcap/{Pcap}
\def\pcapng/{Pcap NG}
\def\keylog/{key-log}
\def\Keylog/{Key-log}
\def\netem/{NetEm}
\def\dummynet/{dummynet}
\def\Nsthree/{\mbox{ns-3}}
\def\nsthree/{\mbox{ns-3}}
\def\Wireshark/{Wireshark}
\def\wireshark/{Wireshark}
\def\Tshark/{TShark}
\def\tshark/{TShark}
\def\Pyshark/{Pyshark}
\def\pyshark/{pyshark}
\def\Tcpdump/{Tcpdump}
\def\tcpdump/{tcpdump}
\def\Qlog/{qlog}
\def\qlog/{qlog}
\def\Qvis/{qvis}
\def\qvis/{qvis}
\def\Python/{Python}
\def\python/{\Python/}
\def\dockerpy/{docker-py}
\def\Docker/{Docker}
\def\docker/{\Docker/}
\def\Doco/{\Docker/ Compose}
\def\doco/{\Doco/}

\def\PC/{\acs{PC}}
\def\IoT/{\acs{IoT}}
\def\URL/{\acs{URL}}

\def\IETF/{\acs{IETF}}
\def\RFC/{\acs{RFC}}

\def\SATCOM/{\acs{SATCOM}}

\def\RTT/{\acs{RTT}}
\def\IP/{\acs{IP}}
\def\ICMP/{\acs{ICMP}}
\def\UDP/{\acs{UDP}}
\def\TCP/{\acs{TCP}}
\def\SCTP/{\acs{SCTP}}
\def\WebRTC/{\acs{WebRTC}}
\def\QUIC/{QUIC}
\def\gQUIC/{gQUIC}
\def\HTTP/{\acs{HTTP}}
\def\TLS/{\acs{TLS}}
\def\LAN/{\acs{LAN}}
\def\VPN/{\acs{VPN}}
\def\DNS/{\acs{DNS}}
\def\MASQUE/{\acs{MASQUE}}

\def\BBR/{\acs{BBR}}
\def\BBRv2/{\BBR/v2}
\def\BIC/{BIC}
\def\CUBIC/{CUBIC}
\def\Reno/{Reno}
\def\NewReno/{NewReno}
\def\pico/{pico}

\def\SYN/{\texttt{SYN}}
\def\SYNs/{\texttt{SYNs}}
\def\ACK/{\texttt{\acs{ACK}}}
\def\ACKs/{\texttt{\acsp{ACK}}}
\def\NACK/{\texttt{NACK}}
\def\FIN/{\texttt{FIN}}
\def\STREAM/{\texttt{STREAM}}

\def\aioquic/{\textit{aioquic}}
\def\chrome/{\textit{chrome}}
\def\kwik/{\textit{kwik}}
\def\lsquic/{\textit{lsquic}}
\def\msquic/{\textit{msquic}}
\def\mvfst/{\textit{mvfst}}
\def\neqo/{\textit{neqo}}
\def\nginx/{\textit{nginx}}
\def\ngtcp/{\textit{ngtcp2}}
\def\picoquic/{\textit{picoquic}}
\def\quant/{\textit{quant}}
\def\quicgo/{\textit{quic-go}}
\def\quiche/{\textit{quiche}}
\def\quicly/{\textit{quicly}}
\def\xquic/{\textit{xquic}}

\def\Aioquic/{\textit{Aioquic}}
\def\Chrome/{\textit{Chrome}}
\def\Kwik/{\textit{Kwik}}
\def\Lsquic/{\textit{Lsquic}}
\def\Msquic/{\textit{Msquic}}
\def\Mvfst/{\textit{Mvfst}}
\def\Neqo/{\textit{Neqo}}
\def\Nginx/{\textit{NGINX}}
\def\Ngtcp/{\textit{Ngtcp2}}
\def\Picoquic/{\textit{Picoquic}}
\def\Quant/{\textit{Quant}}
\def\Quicgo/{\textit{Quic-go}}
\def\Quiche/{\textit{Quiche}}
\def\Quicly/{\textit{Quicly}}
\def\Xquic/{\textit{Xquic}}

\newcommand{\makeunit}[1]{{\mdseries\color{aluminium6}{$[\si{#1}]$}}}

\pgfmathsetseed{\number\pdfrandomseed}

\newboolean{doubleblind}
\setboolean{doubleblind}{false}
\newcommand{\hideurlwhendoubleblind}[1]{%
    \pgfmathsetmacro{\randomurl}{ifthenelse(rnd<0.5,"https://youtu.be/dQw4w9WgXcQ","https://http.cat/401")}%
    \ifthenelse{\boolean{doubleblind}}{%
        \href{\randomurl}{%
            \textit{[Link removed due to double-blind peer review, but will be provided in final version.]}
        }
    }{%
        \url{#1}
    }
}

\makeatletter
\AtBeginDocument{
  \hypersetup{
    pdftitle={\@title},
  }
}
\makeatother

\title{Performance of \QUIC/ Implementations Over Geostationary Satellite Links}

\ifthenelse{\boolean{doubleblind}}{%
    \author{\textit{Authors removed, due to double-blind peer review}}
}{%
    \author{%
        \IEEEauthorblockN{Sebastian Endres, J\"org Deutschmann, Kai-Steffen Hielscher, and Reinhard German}
        \IEEEauthorblockA{%
            \textit{Computer Networks and Communication Systems}\\
            \textit{University of Erlangen-Nuremberg}\\
            \{%
                \href{mailto:basti.endres@fau.de}{basti.endres},
                \href{mailto:joerg.deutschmann@fau.de}{joerg.deutschmann},
                \href{mailto:kai-steffen.hielscher@fau.de}{kai-steffen.hielscher},
                \href{mailto:reinhard.german@fau.de}{reinhard.german}%
            \}@fau.de
        }
    }
}

\begin{document}

\acrodef{ACK}{Acknowledgement}
\acrodef{API}{Application Programming Interface}
\acrodef{AQM}{Active Queue Management}
\acrodef{ARQ}{Automatic Repeat-Request}
\acrodef{AVX}{Advanced Vector Extensions}
\acrodef{BBR}{Bottleneck Bandwidth and Round-trip propagation time}
\acrodef{BDP}{Bandwidth Delay Product}
\acrodef{BER}{Bit Error Ratio}
\acrodef{CCA}{Congestion Control Algorithm}
\acrodef{CDF}{Cumulative Distribution Function}
\acrodef{cnes}{Centre national d’études spatiales}
\acrodef{CPU}{Central Processing Unit}
\acrodef{cwin}{congestion window}
\acrodef{DNS}{Domain Name System}
\acrodef{DPLPMTUD}{Datagram Packetization Layer \acs{PMTUD}}
\acrodef{DupACK}{Duplicate \acl{ACK}}
\acrodef{DVB}{Digital Video Broadcasting}
\acrodef{ECN}{Explicit Congestion Notification}
\acrodef{ELK-stack}{Elasticsearch, Logstash, Kibana}
\acrodef{FEC}{Forward Error Correction}
\acrodef{GEO}{Geostationary Earth Orbit}
\acrodef{GLONASS}{Global Navigation Satellite System}
\acrodef{GPS}{Global Positioning System}
\acrodef{GS}{Ground Station}
\acrodef{GUI}{Graphical User Interface}
\acrodef{HEO}{High Elliptical Earth Orbit}
\acrodef{HoLB}{Head-of-Line-blocking}
\acrodef{HTS}{High Throughput Satellite}
\acrodef{HTTP}{Hypertext Transfer Protocol}
\acrodef{ICMP}{Internet Control Message Protocol}
\acrodef{ID}{Identifier}
\acrodef{IETF}{Internet Engineering Task Force}
\acrodef{IoT}{Internet of Things}
\acrodef{IP}{Internet Protocol}
\acrodef{ISL}{Inter-Satellite Link}
\acrodef{ISP}{Internet Service Provider}
\acrodef{ITU-R}{International Telecommunication Union, Radiocommunication Sector}
\acrodef{iwin}{initial congestion window}
\acrodef{JSON}{JavaScript Object Notation}
\acrodef{KDE}{kernel density estimation}
\acrodef{LAN}{Local Area Network}
\acrodef{LEO}{Low Earth Orbit}
\acrodef{LFN}{Long Fat Network}
\acrodef{MASQUE}{Multiplexed Application Substrate over \QUIC/ Encryption}
\acrodef{MEO}{Medium Earth Orbit}
\acrodef{MobSatServ}[MSS]{Mobile Satellite Services}
\acrodef{MSS}{Maximum Segment Size}
\acrodef{MTU}{Maximum Transmission Unit}
\acrodef{NAT}{Network Address Translation}
\acrodef{NCC}{Network Control Center}
\acrodef{OpenBACH}{Open Benchmark Automation tools for Communication and Hypervision}
\acrodef{OpenSAND}{Open Satellite Network Demonstrator}
\acrodef{PC}{Personal Computer}
\acrodef{PEO}{Polar Earth Orbit}
\acrodef{PEP}{Performance Enhancing Proxy}
\acrodef{PLR}{Packet Loss Rate}
\acrodef{PLT}{Page Load Time}
\acrodef{PMTU}{Path \acs{MTU}}
\acrodef{PMTUD}{Path \acs{MTU} Discovery}
\acrodef{PTO}{Probe Timeout}
\acrodef{QEF}{quasi error-free}
\acrodef{QIR}{\QUIC/-Interop-Runner}
\acrodef{QIR-SE}{\acl{QIR} Satellite Edition}
\acrodef{QNS}{\QUIC/ Network Simulator}
\acrodef{QoE}{Quality of Experience}
\acrodef{QoS}{Quality of Service}
\acrodef{RACK}{Recent Acknowledgement}
\acrodef{RAM}{Random Access Memory}
\acrodef{rcwin}{Receive Window}
\acrodef{RED}{Random Early Detection}
\acrodef{RFC}{Request for Comments}
\acrodef{RTT}{Round-Trip Time}
\acrodef{SACK}{Selective \acl{ACK}}
\acrodef{SATCOM}{satellite communications}
\acrodef{SCC}{Satellite Control Center}
\acrodef{SCTP}{Stream Control Transmission Protocol}
\acrodef{SDG}{Sustainable Development Goal}
\acrodef{SGS}{Source \acl{GS}}
\acrodef{SIMD}{Single Instruction Multiple Data}
\acrodef{SLA}{Service Level Agreement}
\acrodef{SSE}{Streaming \acs{SIMD} Extensions}
\acrodef{SSH}{Secure Shell}
\acrodef{STK}{Satellite Tool Kit}
\acrodef{SWOT}{strengths, weaknesses, opportunities, and threads}
\acrodef{TCP}{Transmission Control Protocol}
\acrodef{TDMA}{Time Division Multiple Access}
\acrodef{TFO}{\TCP/ Fast Open}
\acrodef{TLP}{Tail Loss Probe}
\acrodef{TLS}{Transport Layer Security}
\acrodef{TTC}[TT\&C]{Telemetry, Tracking and Control}
\acrodef{TTFB}{Time to first Byte}
\acrodef{TTLB}{Time to last Byte}
\acrodef{TTR}{Time to \texttt{responseStart}}
\acrodef{UDP}{User Datagram Protocol}
\acrodef{URL}{Uniform Resource Locator}
\acrodef{VLEO}{Very Low Earth Orbit}
\acrodef{VM}{Virtual Machine}
\acrodef{VPN}{Virtual Private Network}
\acrodef{WebRTC}{Web Real-Time Communication}
\acrodefplural{PEP}{Per\-for\-mance En\-han\-cing Proxies}

\maketitle

\begin{abstract}
    \QUIC/ was recently standardized as \RFC/~9000, but the performance of \QUIC/ over geostationary satellite links is problematic due to the non-applicability of \aclp{PEP}.
    As of today, there are more than a dozen of different \QUIC/ implementations.
    So far performance evaluations of \QUIC/ over satellite links were limited to specific \QUIC/ implementations.
    By deploying a modified version of the IETF \acl{QIR}, this paper evaluates the performance of multiple \QUIC/ implementations over multiple geostationary satellite links.
    This includes two emulated ones (with and without packet loss) and two real ones.
    The results show that the goodput achieved with \QUIC/ over geostationary satellite links is very poor in general, and especially poor when there is packet loss.
    Some implementations fail completely and the performance of the other implementations varies greatly.
    The performance depends on both client and server implementation.
\end{abstract}

\section{Introduction}%
\label{sec:introduction}

\begin{figure*}[t]
    \centering
    \includegraphics[width=\linewidth]{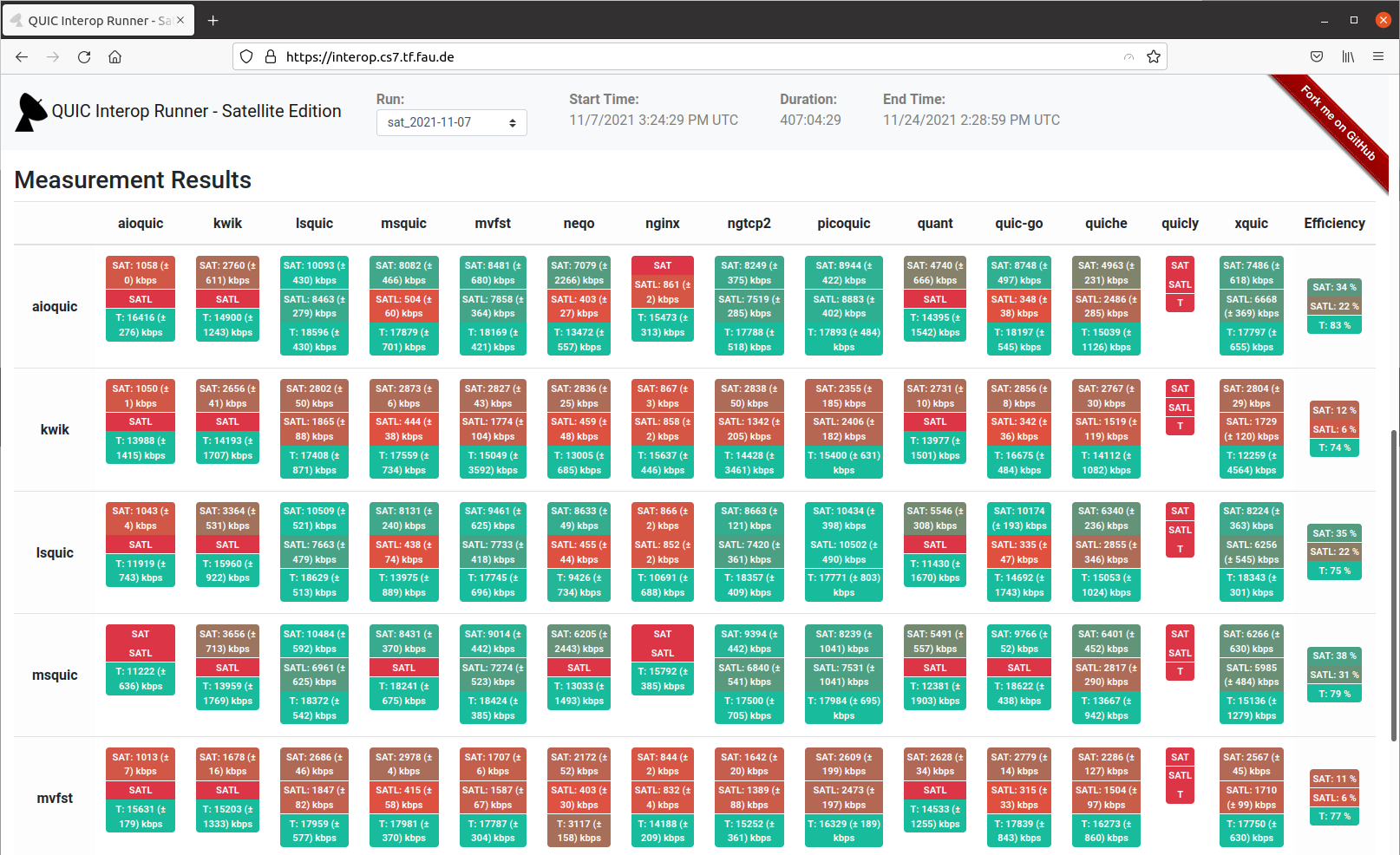}
    \caption{Partial Screenshot of the \ac{QIR-SE} Website Showing the Measurement Results}%
    \label{fig:screenshot}
\end{figure*}

Thanks to their wide-area coverage, satellites are the only option for Internet access for some people.
Today's modern geostationary high throughput satellites (GEO) provide data rates up to \SI{100}{\mbps} in the downlink and even more in the future~\cite{gaudenziFutureTechnologiesVeryHighThroughputSatelliteSystems2020}.
However, compared to other Internet access technologies, geostationary satellites suffer from high propagation delays.
Together with other delays, \acp{RTT} of \SI{600}{\milli\second} and more are typical~\cite{deutschmannSatelliteInternetPerformance2019}.
Obviously, latency-sensitive applications are inherently problematic for GEO satellites, but also the performance of transport protocols like \TCP/ is problematic over high-delay links because of slow-start and long feedback loops.
In order to mitigate poor \TCP/ performance, \acp{PEP} are deployed in satellite networks~\cite{rfc3135}.
With encrypted transport layer headers, as it is the case for \acsp{VPN} and \QUIC/~\cite{rfc9000}, \acp{PEP} can not be applied anymore.
This leads to significant performance degradation described in earlier publications~\cite{%
    secchiEvaluatingPerformanceNext2015,
    thomasGoogleQUICPerformance2019,
    deutschmannSatelliteInternetPerformance2019,
    mogildeaQUICSatelliteIntroduction2019,borderEvaluatingQUICPerformance2020,kuhnQUICOpportunitiesThreats2020,custuraImpactAcknowledgementsUsing2020%
} and this work.
Research, like \cite{%
    yangPerformanceAnalysisQUIC2018,
    zhangHowQuickQUIC2018%
}, that neglects the fact that \acp{PEP} are usually used to accelerate \TCP/ connections, conclude that \QUIC/ outperforms \TCP/ on these satellite links.
This is also confirmed by some of the before mentioned references, which compare \QUIC/'s performance with \TCP/ used over \acsp{VPN}.
We prepared a more detailed work in progress literature survey which is available online\footnote{%
    \hideurlwhendoubleblind{https://github.com/sedrubal/QUIC_HIGH_BDP/blob/master/research_overview.md}
}.

The \ac{QIR}\footnote{%
    \url{https://interop.seemann.io}
}~\cite{seemannAutomatingQUICInteroperability2020} is used by the \IETF/ \QUIC/ Working Group\footnote{%
    \url{https://datatracker.ietf.org/wg/quic} and \url{https://quicwg.org}
} for interoperability testing of \QUIC/ implementations.
It already supports numerous implementations uniformly packaged into \docker/\footnote{%
    \url{https://docker.com}
} containers and available on \docker/ Hub\footnote{%
    \url{https://hub.docker.com}
}.
The runner tests every client implementation with every server implementation three times per day and performs several checks, like if a handshake completes successfully, or if 0-RTT works.
The following implementations, grouped by their role, were part of the \ac{QIR} at the time of writing:
\begin{itemize}
    \item Client \& Server:
          \aioquic/,
          \kwik/,
          \lsquic/,
          \msquic/,
          \mvfst/,
          \neqo/,
          \ngtcp/,
          \picoquic/,
          \quant/,
          \quicgo/,
          \quiche/,
          \quicly/,
          \xquic/
    \item Client only: \chrome/
    \item Server only: \nginx/
\end{itemize}
For a detailed description of the implementations, please refer to the \ac{QIR} website or the \QUIC/ Working Group\footnote{%
    \url{https://github.com/quicwg/base-drafts/wiki/Implementations}
}.
In the official \ac{QIR}, \chrome/ was not supported and \quicly/ always failed.
These two implementations are thus also ignored in our tests.

The focus of the \ac{QIR} is on testing interoperability, functionalities, and standards compliance.
There are also two performance measurement tests: \g/ uses an emulated low-latency channel with a data rate of \SI{10}{\mbps} to measure the average goodput for the transfer of a large file, and \crosstraffic/ measures the goodput with a competing \texttt{iperf}\footnote{%
    \url{https://iperf.fr}
} transmission.
These non-challenging link parameters lead to good results for almost all implementations, i.e., the achieved goodput is close to the physical layer link rate of \SI{10}{\mbps}.
However, real Internet access links, especially geostationary satellite links with high delays, are much more challenging~\cite{jonesEnhancingTransportProtocols2021}.
This motivated us to adapt the \ac{QIR} to include geostationary satellite links, which we then call \ac{QIR-SE}\footnote{%
    \label{fn:qir_se_website}\hideurlwhendoubleblind{https://interop.cs7.tf.fau.de}
}.
To the best of our knowledge, this is the first time that the performance of a broad range of \QUIC/ implementations over geostationary satellite links is evaluated.

The architecture of \ac{QIR} and \ac{QIR-SE} is described in \cref{sec:architecture}.
To gain further insights into realistic use cases, we also conducted tests over real satellite links.
The configuration of the emulated satellite links and the used real satellite accesses are described in \cref{sec:testsetup}.
Similar to \ac{QIR}, all results obtained with the \ac{QIR-SE} are presented on a website\textsuperscript{\ref{fn:qir_se_website}}, which is shown in \cref{fig:screenshot}.
In \cref{sec:evaluation} results are elaborated, which in addition to the original \ac{QIR} also include auto-generated time-offset diagrams.
\cref{sec:conclusion} concludes this paper and suggests future research.

\section{Architecture}%
\label{sec:architecture}

\begin{figure}
    \centering
    \subfloat[Architecture of the Original \ac{QIR} used for Emulations]{%
        \resizebox{\linewidth}{!}{%
            \let\ArchNoCrossTraffic=t
            \def\containerDist/{2.5cm}

\begin{tikzpicture}[
        base/.style={thick, align=center, text centered},
        node/.style={base, minimum height=1cm, minimum width=1.75cm, draw=Orange, fill=LightOrange!55},
        slot/.style={node, draw=Orange, fill=white, loosely dashed},
        largearrowtipStart/.style={decoration={markings, mark=at position 2 with {\arrow[scale=1.5]{<}}}, postaction={decorate}, shorten >=0.4pt},
        largearrowtipEnd/.style={decoration={markings, mark=at position 1 with {\arrow[scale=1.5]{>}}}, postaction={decorate}, shorten >=0.4pt},
        link/.style={largearrowtipStart, largearrowtipEnd, to-to, ultra thick},
        transfer/.style={-to, draw=DarkPlum, color=DarkPlum, thick},
        manage/.style={-to, dotted, draw=ScarletRed, color=ScarletRed, thick},
        host/.style={draw=aluminium5, dashed, inner sep=3mm},
        eth/.style={fill=Orange, anchor=base},
    ]

    \node[node, minimum width=9cm, fill=LightSkyBlue!55, draw=LightSkyBlue] (doco) {\Doco/\phantom{X}};

    \node[node, minimum width=11cm, below=1cm of doco, fill=DarkSkyBlue!55, draw=DarkSkyBlue, text=white] (qir) {\textbf{\QUIC/-Interop-Runner}};
    \node[base, below=10pt of qir.south west, anchor=north west, color=aluminium5] (lblHost) {Host};

    \ifx\ArchNoCrossTraffic t
    \else
        \node[slot, above=\containerDist/+1.25cm of doco.north west, anchor=south west] (client2) {};
    \fi
    \node[node, above=\containerDist/ of doco.north west, anchor=south west] (client) {\textbf{Client}};

    \node[node, above=\containerDist/ of doco.north, anchor=south] (ns3) {\textbf{\phantom{C}\nsthree/\phantom{C}}};

    \ifx\ArchNoCrossTraffic t
    \else
        \node[slot, above=\containerDist/+1.25cm of doco.north east, anchor=south east] (server2) {};
    \fi
    \node[node, above=\containerDist/ of doco.north east, anchor=south east] (server) {\textbf{Server}};

    \begin{scope} 
        \node[eth] (clientEth) at (client.east) {};
        \node[eth] (ns3LeftEth) at (ns3.west) {};
        \node[eth] (ns3RightEth) at (ns3.east) {};
        \node[eth] (serverEth) at (server.west) {};
        \ifx\ArchNoCrossTraffic t
        \else
            \node[eth] (client2Eth) at (client2.east) {};
            \node[eth] (server2Eth) at (server2.west) {};
        \fi
    \end{scope}

    \vertCurledArrow{link}{1cm}{clientEth}{ns3LeftEth};
    \vertCurledArrow{link}{1cm}{ns3RightEth}{serverEth};
    \ifx\ArchNoCrossTraffic t
    \else
        \vertCurledArrow{link,loosely dashed}{1cm}{client2Eth}{ns3LeftEth};
        \vertCurledArrow{link,loosely dashed}{1cm}{ns3RightEth}{server2Eth};
    \fi

    \ifx\ArchNoCrossTraffic t
        \node[fit=(client) (ns3) (server)] (containers) {};
    \else
        \node[fit=(client) (ns3) (server) (client2) (server2)] (containers) {};
    \fi

    \begin{scope}
        \draw[manage] ([xshift=-5mm]intersection of doco.north west--doco.north east and client.north--client.south) -- ([xshift=-5mm]client.south)
        node[midway,sloped,above] {deploy};
        \draw[manage] ([xshift=-5mm]doco.north) -- ([xshift=-5mm]ns3.south)
        node[midway,sloped,above] {deploy};
        \draw[manage] ([xshift=-5mm]intersection of doco.north west--doco.north east and server.north--server.south) -- ([xshift=-5mm]server.south)
        node[midway,sloped,below] {deploy};

        \draw[manage] ([xshift=-5mm]qir.north) -- ([xshift=-5mm]doco.south)
        node[midway,sloped,above] {run};
    \end{scope}

    \begin{scope}[on background layer]
        \draw[transfer] ([xshift=+5mm]client.south) -- ([xshift=+5mm]intersection of qir.north west--qir.north east and client.north--client.south)
        node[midway,pos=0.3,sloped,below] {qlog, keylog};
        \draw[transfer] ([xshift=+5mm]ns3.south) -- ([xshift=+5mm]qir.north)
        node[midway,pos=0.3,sloped,above] {pcap};
        \draw[transfer] ([xshift=+5mm]server.south) -- ([xshift=+5mm]intersection of qir.north west--qir.north east and server.north--server.south)
        node[midway,pos=0.3,sloped,above] {qlog, keylog};
    \end{scope}

    \node[host, fit=(containers) (qir) (lblHost.center)] {};

\end{tikzpicture}
        }
        \label{fig:qir_architecture_original}
    }

    \subfloat[Distributed Architecture of \ac{QIR} used for Real Satellite Links]{%
        \resizebox{\linewidth}{!}{%
            \let\ArchIncreaseFont=t
            \def\mgmtYShift/{2.5mm}
\def\mgmtXShift/{5mm}
\def\innerEthDist/{5mm}
\def\hostLblDist/{5mm}
\def\smallNodeWidth/{22mm}
\def\largeNodeWidth/{2*\smallNodeWidth/}

\begin{tikzpicture}[
        base/.style={thick, align=center, text centered},
        node/.style={
                base,
                minimum height=1cm,
                minimum width=\smallNodeWidth/,
                draw=Orange,
                fill=LightOrange!55,
                font={\ifx\ArchIncreaseFont t\large\fi},
            },
        container/.style={node},
        docker/.style={node, fill=LightSkyBlue!55, draw=LightSkyBlue},
        largearrowtipStart/.style={decoration={markings, mark=at position 2 with {\arrow[scale=1.5]{<}}}, postaction={decorate}, shorten >=0.4pt},
        largearrowtipEnd/.style={decoration={markings, mark=at position 1 with {\arrow[scale=1.5]{>}}}, postaction={decorate}, shorten >=0.4pt},
        link/.style={largearrowtipStart, largearrowtipEnd, to-to, ultra thick},
        transfer/.style={-to, color=DarkPlum, thick},
        manage/.style={-to, dotted, color=ScarletRed, thick},
        host/.style={draw=aluminium5, dashed, inner sep=\innerEthDist/},
        hostLbl/.style={
                color=aluminium5,
                font={\ifx\ArchIncreaseFont t\large\fi},
            },
        eth/.style={fill=Orange, anchor=base},
        mgmtLink/.style={largearrowtipStart, largearrowtipEnd, to-to, ultra thick, color=ScarletRed},
        nodeBetween/.style={node, draw=Chocolate, fill=LightChocolate!55},
        sat/.style={draw=Chocolate, fill=LightChocolate!55},
    ]

    \begin{scope} 
        \node[nodeBetween, draw=none, fill=none] (satellite) {};
        \satellite{(satellite.center)}{90}{0.8};

        \node[nodeBetween, below left=5mm and 2mm of satellite.south] (modem) {Modem};
        \node[nodeBetween, below right=5mm and 2mm of satellite.south] (internet) {Internet};

        \draw[link] (modem) -- ([xshift=-1mm]satellite.south);
        \draw[link] ([xshift=1mm]satellite.south) -- (internet);

        \node[color=DarkChocolate,yshift=-\hostLblDist/] at (satellite.center |- modem.south) (linkLbl) {Link};
        \node[node, draw=DarkChocolate, fill=none, dashed, inner sep=\innerEthDist/, fit=(modem) (satellite) (internet) (linkLbl.center)] (link) {};
    \end{scope}

    \begin{scope} 
        \node[eth,right=\innerEthDist/ of internet] (ethSrvHost) {};
        \node[eth,right=\innerEthDist/-2mm of ethSrvHost] (ethSrv) {};

        \begin{scope}[on background layer]
            \node[container, above right=0 and 0 of ethSrv.center, anchor=south west] (server) {\textbf{Server}};
            \node[container, below right=0 and 0 of ethSrv.center, anchor=north west] (tcpdumpSrv) {\ifx\ArchIncreaseFont t\else\small\fi\Tcpdump/};
        \end{scope}

        \node[node, below=3.75cm of ethSrv.center, anchor=west, minimum width=\largeNodeWidth/, fill=DarkSkyBlue!55, draw=DarkSkyBlue, text=white] (qir) {\textbf{\QUIC/-Interop-Runner}};
        \node[docker, above right=15mm and 0 of qir.east, anchor=east] (dockerSrv) {\Docker/};
        \node[eth, anchor=east, left=\innerEthDist/ of qir.west] (ethSrvMgmt) {};

        \draw[ultra thick] (ethSrvHost) -- (ethSrv);
        \draw[mgmtLink] (qir) -- (ethSrvMgmt);

        \draw[manage] ([xshift=-\mgmtXShift/]qir.north east) -- ([xshift=-\mgmtXShift/]dockerSrv.south east)
        node[midway,right] {call};
        \draw[manage] ([xshift=-\mgmtXShift/]dockerSrv.north east) |- ([yshift=\mgmtYShift/]tcpdumpSrv.east);
        \draw[manage] ([xshift=-\mgmtXShift/]dockerSrv.north east) |- ([yshift=\mgmtYShift/]server.east)
        node[midway,sloped,pos=0.3,fill=white] {deploy};

        \begin{scope}[on background layer]
            \draw[transfer] ([yshift=-\mgmtYShift/]tcpdumpSrv.east) -| ([xshift=7mm]qir.north);
            \draw[transfer] ([yshift=-\mgmtYShift/]server.east) -| ([xshift=7mm]qir.north)
            node[midway,left,pos=0.97] {pcap, qlog, keylog};
        \end{scope}

        \node[hostLbl, below left=\hostLblDist/ and 0 of qir.south east, anchor=north east] (hostSrvLbl) {Server Host};
        \node[host, fit=(hostSrvLbl.center) (qir) (server)] (hostSrv) {};
    \end{scope}

    \begin{scope} 
        \node[eth, left=\innerEthDist/ of modem] (ethCliHost) {};
        \node[eth, left=\innerEthDist/-2mm of ethCliHost] (ethCli) {};

        \begin{scope}[on background layer]
            \node[container, above left=0 and 0 of ethCli.center, anchor=south east] (client) {\textbf{Client}};
            \node[container, below left=0 and 0 of ethCli.center, anchor=north east] (tcpdumpCli) {\ifx\ArchIncreaseFont t\else\small\fi\Tcpdump/};
        \end{scope}

        \node[node, below=3.75cm of ethCli.center, anchor=east, fill=LightPlum!55, draw=LightPlum] (ssh) {SSH};
        \node[docker, above left=15mm and 0 of ssh.west, anchor=east] (dockerCli) {\Docker/};
        \node[eth, anchor=west, right=\innerEthDist/ of ssh.east] (ethCliMgmt) {};

        \draw[mgmtLink] (dockerCli) -| (ssh)
        node[midway] (vertCenterCli) {};
        \draw[ultra thick] (ethCliHost) -- (ethCli);
        \draw[mgmtLink] (ssh) -- (ethCliMgmt);

        \draw[manage] ([xshift=\mgmtXShift/]dockerCli.north west) |- ([yshift=\mgmtYShift/]tcpdumpCli.west);
        \draw[manage] ([xshift=\mgmtXShift/]dockerCli.north west) |- ([yshift=\mgmtYShift/]client.west)
        node[midway,sloped,pos=0.3,fill=white] {deploy};

        \begin{scope}[on background layer]
            \draw[transfer] ([yshift=-\mgmtYShift/]tcpdumpCli.west) -| (dockerCli.north);
            \draw[transfer] ([yshift=-\mgmtYShift/]client.west) -| (dockerCli.north)
            node[midway,right,pos=0.9] {logs};
        \end{scope}

        \node[hostLbl, anchor=north west] (hostCliLbl) at (dockerCli.west |- hostSrvLbl.north) {Client Host};
        \node[host, fit=(hostCliLbl.center) (dockerCli) (ssh) (client) (tcpdumpCli)] (hostCli) {};
    \end{scope}

    \draw[link] (ethCliHost) -- (modem);
    \draw[link] (internet) -- (ethSrvHost);
    \draw[mgmtLink] (ethCliMgmt) -- (ethSrvMgmt)
    node[midway, below] {control};

\end{tikzpicture}
        }
        \label{fig:qir_architecture_distributed}
    }

    \caption{Architecture of The \acs{QIR} \& \acs{QIR-SE}}%
    \label{fig:qir_architecture}
\end{figure}
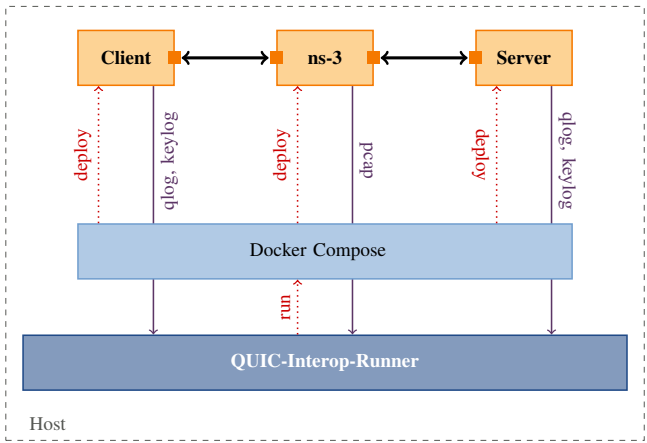
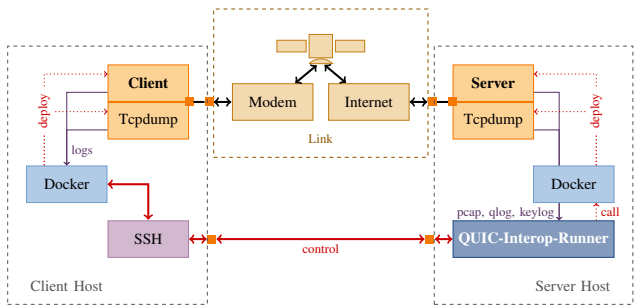

The architecture of \ac{QIR} is shown in \cref{fig:qir_architecture_original}.
It employs \doco/\footnote{%
    \Doco/:
    \url{https://docs.docker.com/compose}
} with virtual networks to deploy the entire test environment on a single host machine.
Most test cases require a setup of three containers:
The \QUIC/ client implementation, the \QUIC/ server implementation and a container containing the \nsthree/ Network Simulator\footnote{%
    \url{https://nsnam.org}
} in between.
Additional containers can be deployed for special test cases.
For the link emulator, different \nsthree/ scenarios are used, which are available online\footnote{%
    Repository of \nsthree/ scenarios:
    \url{https://github.com/marten-seemann/quic-network-simulator}
}.
The runner passes \TLS/ keys and certificates~\cite{rfc8446} through the \docker/ volume \texttt{/certs} into the end point containers.
The implementations have to generate log files in the \docker/ volume \texttt{/logs}.
Servers have to serve files with random content from the volume \texttt{/www} and clients have to download them to the volume \texttt{/downloads}, preferably by using \HTTP//0.9, which is very simple and has a minimal overhead\footnote{%
    Discussion about using \HTTP//0.9 as protocol:
    \url{https://github.com/marten-seemann/quic-interop-runner/issues/267}
}.
During the run, \nsthree/ is used to capture packets on the wire into \pcap/ files.
The traces are then analyzed using \pyshark/\footnote{%
    \url{https://kiminewt.github.io/pyshark}
}, a \python/ wrapper for \texttt{tshark} from the \wireshark/ protocol analyzer\footnote{%
    \url{https://wireshark.org}
} tool set.
As each client implementation is tested with each server implementation, a two-dimensional matrix is filled with test results, which is visualized in \cref{fig:screenshot,fig:result_matrices}.
It is stored as \acs{JSON} file.
We use this information and the captured traces to automatically generate time-offset plots for each combination of implementation afterwards, as shown in \cref{sec:detailed_graphs}.

\subsection{Emulation}%
\label{sec:architecture_emulation}

To emulate satellite links, which have usually different properties in the forward and return path, we added a new asymmetric ns-3 scenario\footnote{%
    \hideurlwhendoubleblind{https://github.com/sedrubal/quic-network-simulator/tree/feature-asymmetric-p2p/sim/scenarios/asymmetric-p2p}
}.
The data rate, emulated queue size and \ac{PLR} can be configured individually per direction.
The artificial delay applies to both directions.

\subsection{Real Satellite Links}%
\label{sec:architecture_real}

Additionally, we modified the runner to use real network links between the implementations under test.
We replaced the deployment mechanism, which was built upon \doco/, with direct calls to the \docker/ \acs{API} to possibly distributed \docker/ deamons, as visualized in \cref{fig:qir_architecture_distributed}.
In our setup, we have one dedicated machine per satellite access.
\ac{QIR-SE} picks the according machine per test case to deploy the client.
The server is executed on the same machine as the \ac{QIR-SE} itself.
As ns-3 can no longer be used to capture the \pcap/ traces, we run a \texttt{tcpdump} container per end point, which joins the network of the implementation container on the correspondent host.

\section{Test Setup}%
\label{sec:testsetup}

All measurements have been performed on the following systems:

The emulated scenarios have been executed on an Ubuntu 20.04.3 LTS server with Kernel 5.4.0 and \Docker/ 20.10.10.
The server is powered by an \emph{Intel\textregistered\ Xeon\textregistered\ X5650} \acs{CPU} and has \SI{16}{\gib} of \acs{RAM} installed.

For measurements with real satellite connections, the server implementations have been executed on the aforementioned machine.
The clients have been run on other computers, which are connected to the modems of the satellite link.
These computers run Ubuntu 18.04.6 LTS with Kernel 5.4.0 and \Docker/ 20.10.7.
Both computers are powered by an Intel\textregistered\ Core\texttrademark\ i5 4590 \acs{CPU} and \SI{8}{\gib} \acs{RAM} each.

\subsection{Emulation}%
\label{sec:setup_emulation}

We use the \terr/ scenario, which equals the \g/ scenario in the original \ac{QIR}, but with an increased data rate of \SI{20}{\mbps}, to compare the satellite scenarios with a terrestrial link.
The \RTT/ is \SI{30}{\milli\second} and the queue size is 25 packets.
Additionally, we added two performance measurement tests with path properties similar to satellite links:
\sat/ and \satl/.
They differ in the artificial \ac{PLR}, which is set to \SI{0}{\percent} in the first scenario and to \SI{1}{\percent} in the latter one.
In both cases, we set the data rate to 20/\SI{2}{\mbps} (forward / return link) and the one-way delay to \SI{300}{\milli\second}, which results in an \acs{RTT} of \SI{600}{\milli\second}.
All emulated test cases have been repeated ten times per combination of implementation.

\subsection{Real Satellite Links}%
\label{sec:setup_real}

We also ran measurements using two real end-user satellite accesses: \astra/ and \eut/, which are available in Europe.
\ifthenelse{\boolean{doubleblind}}{%
    The satellite dishes and terminals are located \textit{[address removed due to double-blind peer review].}
}{%
    The satellite dishes and terminals are located at the University of Erlangen-Nürnberg, Martensstr. 3, Erlangen, Germany.
}
Details about the tariffs are listed in \cref{tab:real}.
The first one advertises a data rate less than half of the data rate of the second one.
While \eut/ prioritizes traffic up to \SI{60}{\gb}, we did not run into that limit during the measurements.
Both operators provide IPv4 connectivity through carrier-grade \acs{NAT}, and IPv6 is not available.
Further testimonials of the products can be found in a previous measurement campaign\footnote{%
    \label{fn:qos_research_proj}\hideurlwhendoubleblind{https://www.cs7.tf.fau.eu/research/quality-of-service/qos-research-projects/sat-internet-performance}
}, where measurements assessed a stable performance of the real satellite links, and the throughput achieves the advertised link rates.
To be on the safe side, our measurements were carried out during constantly good weather conditions and paused between potentially congested times (i.e., between 6 p.m. and 11 p.m. local time).
Within one run, the execution of client and server combinations was shuffled, and each combination was repeated five times.

\begin{table}
    \centering
    \caption{Measurements with Real Satellite Access}%
    \label{tab:real}
    \begin{tabular}{ccccc}
        \toprule
        \thead{Test                              \\Case}                    &   %
        \thead{Provider                          \\\& Tariff}                      & %
        \thead{Advertised                        \\Data Rate}                       & %
        \thead{Traffic                           \\Limit}                     & %
        \thead{Modem}                            \\
        \midrule
        \makecell{\astra/}                     & %
        \makecell{\textbf{Novostream}            \\Astra Connect\\L+} & %
        \makecell{20/\SI{2}{\mbps}}            & %
        \makecell{\textcolor{aluminium3}{---}} & %
        \makecell{Gilat                          \\SkyEdge\\II-c}\\
        \makecell{\eut/}                       & %
        \makecell{\textbf{Konnect}               \\Zen}            &   %
        \makecell{50/\SI{5}{\mbps}}            & %
        \makecell{\color{aluminium5}priorit.     \\\color{aluminium5}up to\\\SI{60}{\gb}} & %
        \makecell{Hughes                         \\HT2000W}\\
        \bottomrule
    \end{tabular}
\end{table}

\section{Evaluation}%
\label{sec:evaluation}

Before we start with the evaluation, we would like to emphasize that not all implementations might strive for high performance.
Some might be only proof of concept implementations, others might have simplicity or resource efficiency as primary goal, and others might not have been optimized for high latency links yet.
Yet, the performance degradation compared to the emulated \terr/ scenario will be clearly visible.

The evaluation is split into four parts.
\hyperref[sec:overview]{First}, we present heatmaps inspired by the \ac{QIR-SE} website and give an overview considering all results, \hyperref[sec:client_vs_server]{followed} by a discussion whether client or server implementation influences the outcomes more.
\hyperref[sec:impact_cca]{Next}, we try to analyze the impact of different \acp{CCA}.
\hyperref[sec:detailed_graphs]{Lastly}, we pick specific combinations and present their behavior by means of time-offset graphs.

\subsection{Overview}%
\label{sec:overview}

\begin{figure*}[!t]
    \centering
    \subfloat[Average Goodput of Measurement \sat/]{%
        \resizebox{0.48\textwidth}{!}{%
            \input{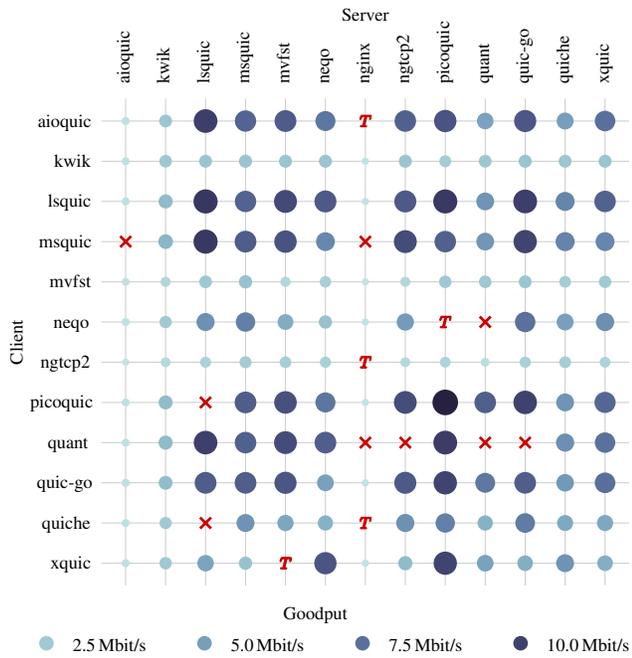}
        }
        \label{fig:result_matrix_sat}
    }
    \hfil
    \subfloat[Average Goodput of Measurement \satl/]{%
        \resizebox{0.48\textwidth}{!}{%
            \input{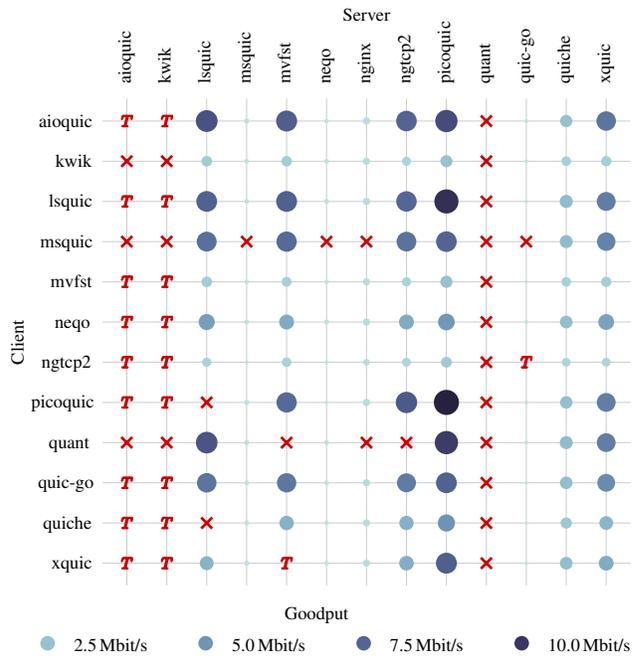}
        }
        \label{fig:result_matrix_satloss}
    }
    \vspace*{0.4cm} %
    \subfloat[Average Goodput of Measurement \astra/]{%
        \resizebox{0.48\textwidth}{!}{%
            \input{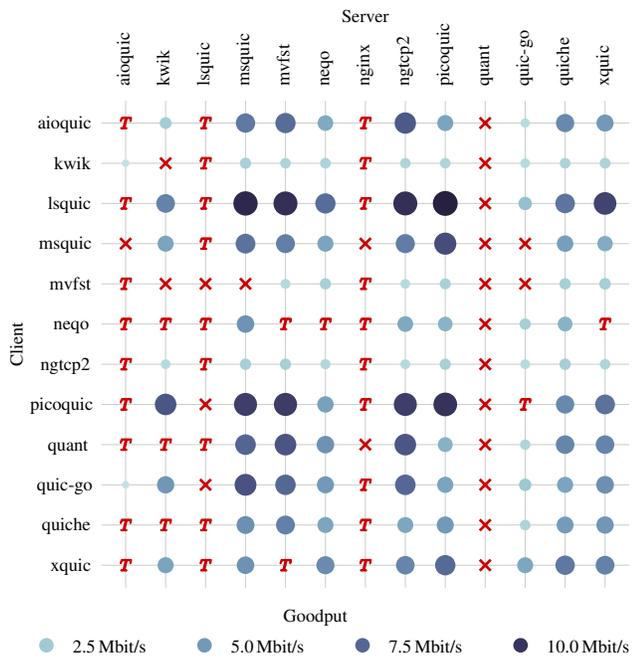}
        }
        \label{fig:result_matrix_ast}
    }
    \hfil
    \subfloat[Average Goodput of Measurement \eut/]{%
        \resizebox{0.48\textwidth}{!}{%
            \input{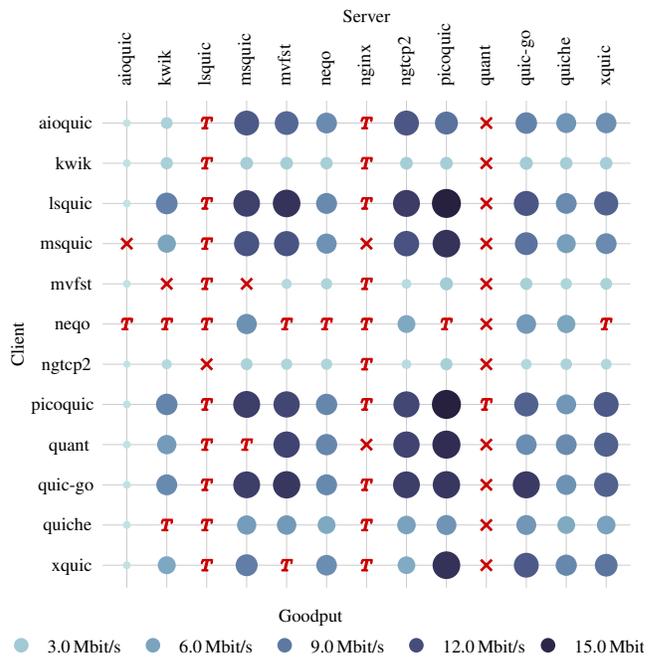}
        }
        \label{fig:result_matrix_eut}
    }
    \vspace*{0.8cm} %
    \caption{Result Matrices (\textcolor{ScarletRed}{\faTimes}: Unspecified Error, \textcolor{ScarletRed}{\boldmath$T$}: Timeout)}
    \label{fig:result_matrices}
\end{figure*}

To visualize the result matrices of the measurements generated with the \ac{QIR-SE} in more detail, we rendered a heatmap for each scenario (\cref{fig:result_matrices}).
The columns belong to the server implementation and the rows to the clients.
We omitted non-functioning implementations.
The color scale is normalized to the minimum and maximum observed goodput value of each scenario.
The size of the circles are synchronized between all charts.
Currently, we only distinguish between unspecified errors (marked with \textcolor{ScarletRed}{\faTimes}) and timeouts (marked with \textcolor{ScarletRed}{\boldmath$T$}).

\begin{table}[t]
    \centering
    \caption[Overview of Measurement Results]{
        Overview of Measurement Results\\
        For the columns \textbf{Mean} and \textbf{Maximum}, the absolute goodput values are in the left column and the relative efficiency values in the right column.
    }%
    \label{tab:overall_stats}
    \setlength\tabcolsep{3.75pt} %
    {%
\begin{tabular}{ccccccc}
    \toprule
    \thead{Measurement} %
       & \multicolumn{2}{c}{\thead{Mean}} %
       & 
         \multicolumn{2}{c}{\thead{Maximum}} %
       & 
         \thead{Timeout \textcolor{ScarletRed}{\boldmath$T$}} %
       & \thead{Failed \textcolor{ScarletRed}{\faTimes}}  \\
       & \makecell{\makeunit{\mbps}} %
       & \makecell{\makeunit{\percent}} %
       & \makecell{\makeunit{\mbps}} %
       & \makecell{\makeunit{\percent}} %
       & \makecell{\makeunit{\percent}} %
       & \makecell{\makeunit{\percent}} \\
    \midrule
    \makecell{\terr/} %
     & \cellcolor{Chameleon!40}15.11 %
     & \cellcolor{Chameleon!40}76 %
     & \cellcolor{Chameleon!40}19.2 %
     & \cellcolor{Chameleon!40}96 %
     & \cellcolor{Orange!5}12 %
     & \cellcolor{Chameleon!20}1 \\
    \makecell{\sat/} %
     & \cellcolor{ScarletRed!20}5.05 %
     & \cellcolor{LightScarletRed!15}25 %
     & \cellcolor{ScarletRed!40}12.0 %
     & \cellcolor{LightOrange!10}60 %
     & \cellcolor{LightChameleon!15}3 %
     & \cellcolor{LightOrange!10}6 \\
    \makecell{\satl/} %
     & \cellcolor{ScarletRed!40}3.06 %
     & \cellcolor{ScarletRed!20}15 %
     & \cellcolor{ScarletRed!40}11.5 %
     & \cellcolor{Orange!5}57 %
     & \cellcolor{DarkOrange!10}13 %
     & \cellcolor{ScarletRed!40}17 \\
    \makecell{\astra/} %
     & \cellcolor{ScarletRed!20}4.91 %
     & \cellcolor{LightScarletRed!15}25 %
     & \cellcolor{LightScarletRed!15}13.5 %
     & \cellcolor{LightOrange!10}68 %
     & \cellcolor{ScarletRed!40}23 %
     & \cellcolor{ScarletRed!20}15 \\
    \makecell{\eut/} %
     & \cellcolor{LightScarletRed!15}6.98 %
     & \cellcolor{ScarletRed!20}14 %
     & \cellcolor{LightChameleon!15}17.5 %
     & \cellcolor{LightScarletRed!15}35 %
     & \cellcolor{ScarletRed!20}20 %
     & \cellcolor{LightScarletRed!15}11 \\
    \arrayrulecolor{black}\bottomrule
\end{tabular}
}

\end{table}

\begin{figure}[t]
    \centering
    \resizebox{\linewidth}{!}{%
        \input{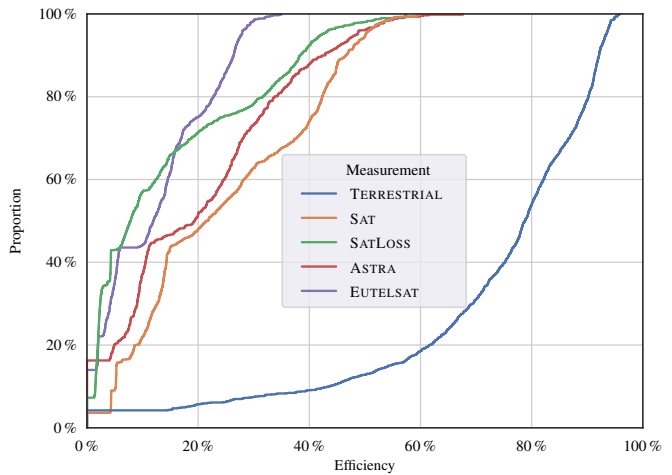}
    }
    \caption{\acsp{CDF} Using Normalized Efficiency Values as Defined in \Cref{eqn:efficiency}}
    \label{fig:cdf_efficiency}
\end{figure}

In addition to the heatmaps, we gathered the measurement results in \cref{tab:overall_stats} and as \acp{CDF} in \cref{fig:cdf_efficiency}.
For each scenario, the table shows the mean and maximum value of all experiments and the amount of combinations that either ran into a timeout or failed for an unknown reason.
On the left side of each column, the absolute goodput values are used (as in the heatmaps).
On the right side, as well as in the \ac{CDF}, we use the \emph{efficiency}, which is defined as the goodput value divided by the emulated or advertised link data rate:

\begin{equation}
    \text{Efficiency} = \frac{\text{Goodput}}{\text{Link Data Rate}}
    \label{eqn:efficiency}
\end{equation}

Besides the emulated satellite scenarios \sat/ and \satl/ and the real satellite links \astra/ and \eut/, the table and the \ac{CDF} plot also contain the results for the \terr/ scenario as reference.
Unsuccessful runs are not considered in the calculation of the mean values in \cref{tab:overall_stats} but plotted as \SI{0}{\percent} efficiency in the \ac{CDF} in \cref{fig:cdf_efficiency}.

First, we take a look at the high number of failed experiments for all scenarios.
The number of failures and timeouts is rather high in the \terr/ scenario because of \neqo/, which did not work as a client in this scenario for unknown reasons, although it worked in our satellite scenarios.
The reason \lsquic/ fails as a server when measuring with real satellite links is technical and caused by the \docker/ image provided by the maintainers.
In the \satl/, \astra/ and \eut/ scenarios we can see a lot of implementations, that were not able to transfer the test file of \SI{10}{\mib} via the corresponding links.
A more detailed analysis of the reasons of failures is subject to future work.

As one can see in \cref{tab:overall_stats} and \cref{fig:cdf_efficiency}, the \terr/ scenario achieves goodput values close to the link rate.
This is not the case for geostationary satellite scenarios.
Instead, there are significant differences between implementations.

Although \sat/, shown in \cref{fig:result_matrix_sat}, has stable link emulation and no artificial loss, there is a lot of variance in the achieved goodput.
Some implementations perform significantly better than the others.
Yet, the achieved goodput is far below the advertised link rate of \SI{20}{\mbps}.
In half of the measurements, it is even below \SI{25}{\percent} of the link data rate.
When all entries of a row (respectively column) show poor results, the implementation performs poor as client (respectively server).
A more detailed analysis of the client vs. server performance of the \sat/ scenario is given in \cref{sec:client_vs_server}.
\satl/ performs worst among all scenarios when considering the absolute goodput values, see \cref{tab:overall_stats}.
\Cref{fig:result_matrix_satloss} shows which implementations suffer most from the \SI{1}{\percent} artificial \ac{PLR} added to the emulated satellite link\footnote{%
    With \QUIC/, a packet loss on any path segment will impact the end to end transmission, while \acp{PEP} are able to retransmit \TCP/ segments locally.
}.
Compared to the no-loss scenario, many implementations do either not finish a transmission successfully or achieve considerably lower goodput rates.
The efficiency decreases from only \SI{25}{\percent} in the \sat/ scenario to even less in the \satl/ scenario (\SI{15}{\percent}).
There are only a few combinations that perform well in the \satl/ scenario, with \picoquic/ as client and server for example being more thoroughly analyzed later in \cref{sec:detailed_graphs_picoquic_picoquic_satl}.

The performance of the real satellite operators is visualized in more breadth in \cref{fig:result_matrix_ast} (\astra/) and \cref{fig:result_matrix_eut} (\eut/).
The \ac{PLR} on these real satellite links is usually very low\textsuperscript{\ref{fn:qos_research_proj}}, but higher dynamics can be expected on real links.
Compared to the \sat/ scenario, this results in less combinations finishing successfully and lower goodput values.
There is a correlation between the results for \astra/ and \eut/, i.e., most combinations perform similarly on both systems.
\eut/ reaches the highest mean and maximum goodput rates in our tests (note different scale in \cref{fig:result_matrix_eut}).
However, given the forward link data rates (\eut/ \SI{50}{\mbps} vs. other satellite scenarios with \SI{20}{\mbps}), the efficiency of \eut/ is with about \SI{14}{\percent} on average actually worse than \astra/ (\SI{25}{\percent}).
This shows that more bandwidth does not automatically result in a better performance by the same factor.
The link utilization on both real satellite links is very low, similar to the results of our simulated test cases.

\subsection{Client vs.\ Server Implementation}%
\label{sec:client_vs_server}

\begin{figure}
    \centering
    \resizebox{\linewidth}{!}{%
        \input{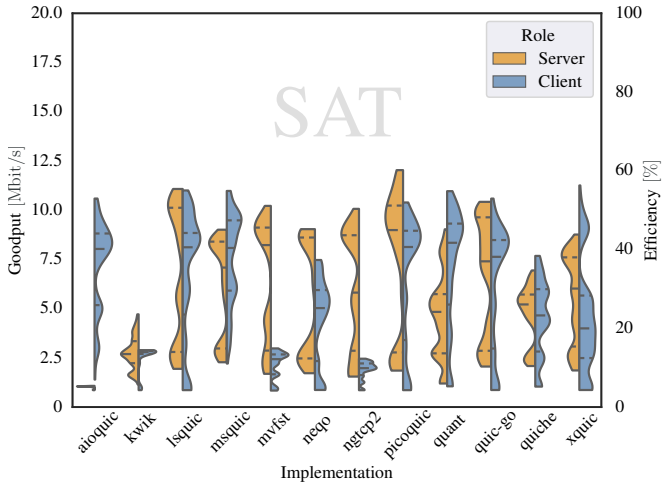}
    }
    \caption{Violin Plots for Each Implementation in the \sat/ Scenario. Horizontal Lines represent the Quartiles $Q_1$, $Q_2$ and $Q_3$.}%
    \label{fig:violins}
    \vspace{0.5em}%
\end{figure}

As the server has to send most of the data and therefore has to estimate the channel parameters and especially the bottleneck bandwidth, it seems likely that the performance of the transmission is mainly determined by the server.
However, the result matrices in \cref{fig:result_matrices} show no pattern that confirms this assumption, instead both the client and the server contribute to the overall performance of a \QUIC/ connection.

In order to verify this statement, we draw violin plots\footnote{%
    Seaborn Documentation for violin plots that employ \ac{KDE}.
    \url{https://seaborn.pydata.org/generated/seaborn.violinplot.html}
} as shown in \cref{fig:violins}.
Due to space limitations, we only present the plot for the \sat/ scenario.
Other scenarios lead to similar results.
The \textcolor{LightOrange}{\faCircle} orange graphs show the performance of a specific server implementation tested with all clients, and the \textcolor{LightSkyBlue}{\faCircle} blue graph shows the performance of a specific client implementation tested with all servers.

Again, it can be seen that the achieved goodput is far off the emulated link rate.
It also becomes clearer, that \kwik/ does neither perform well as client nor as server, \aioquic/ performs well if used as client but not as server, and \mvfst/ and \ngtcp/ perform well if used as server but not as client.
An interesting observation from \cref{fig:violins} is that results are often concentrated towards one end of the distribution.
Especially when looking at the server role, this means that these server implementations either result in very good or very poor results.
Thus, we conclude that both the client and the server contribute to the overall performance of a \QUIC/ connection.

\subsection{Impact of the \acl{CCA}}%
\label{sec:impact_cca}

\begin{table}[t]
    \centering
    \caption{%
        Implementations and their \acsp{CCA}.\\
        Algorithms are determined by hand and must be taken with caution.
        The default \acs{CCA} is printed \textbf{bold}.
    }%
    \label{tab:impls_with_cca}
    {%
\renewcommand{\arraystretch}{1.1}%
\begin{tabular}{ccc}
    \toprule
    \thead{Name} %
       & \thead{\acs{CCA}} %
       & \thead{HyStart} \\
    \midrule
    \makecell{\aioquic/} %
     & \makecell{\textbf{\NewReno/}} %
     & \makecell{\color{aluminium3}\faTimes} \\
    \makecell{\chrome/} %
     & \makecell{\BBRv2/, \CUBIC/} %
     & \makecell{\faCheck} \\
    \makecell{\kwik/} %
     & \makecell{\textbf{\NewReno/}} %
     & \makecell{\color{aluminium3}\faTimes} \\
    \makecell{\lsquic/} %
     & \makecell{\textbf{\BBR/}, \CUBIC/} %
     & \makecell{\color{aluminium3}\faTimes} \\
    \makecell{\msquic/} %
     & \makecell{\textbf{\CUBIC/}} %
     & \makecell{\color{aluminium3}\faTimes} \\
    \makecell{\mvfst/} %
     & \makecell{\BBR/, \textbf{\CUBIC/}, \NewReno/, \textellipsis} %
     & \makecell{\faCheck} \\
    \makecell{\neqo/} %
     & \makecell{\CUBIC/, \textbf{\NewReno/}} %
     & \makecell{\color{aluminium3}\faTimes} \\
    \makecell{\nginx/} %
     & \makecell{\color{aluminium5}\textcircled{?}} %
     & \makecell{\color{aluminium3}\faTimes} \\
    \makecell{\ngtcp/} %
     & \makecell{\BBRv2/, \BBR/, \textbf{\CUBIC/}, \Reno/} %
     & \makecell{\color{aluminium3}\faTimes} \\
    \makecell{\picoquic/} %
     & \makecell{\textbf{\BBR/}, \CUBIC/} %
     & \makecell{\faCheck} \\
    \makecell{\quant/} %
     & \makecell{\textbf{\NewReno/}} %
     & \makecell{\color{aluminium3}\faTimes} \\
    \makecell{\quicgo/} %
     & \makecell{\color{aluminium5}\CUBIC/ \textcircled{?}, \Reno/ \textcircled{?}} %
     & \makecell{\color{aluminium3}\faTimes} \\
    \makecell{\quiche/} %
     & \makecell{\textbf{\CUBIC/}} %
     & \makecell{\faCheck} \\
    \makecell{\quicly/} %
     & \makecell{\CUBIC/, \textbf{\Reno/}, \pico/} %
     & \makecell{\color{aluminium3}\faTimes} \\
    \makecell{\xquic/} %
     & \makecell{\textbf{\BBR/}, \CUBIC/, \Reno/} %
     & \makecell{\color{aluminium3}\faTimes} \\
    \arrayrulecolor{black}\bottomrule
\end{tabular}
}

\end{table}

\begin{figure}[t]
    \centering
    \resizebox{\linewidth}{!}{%
        \input{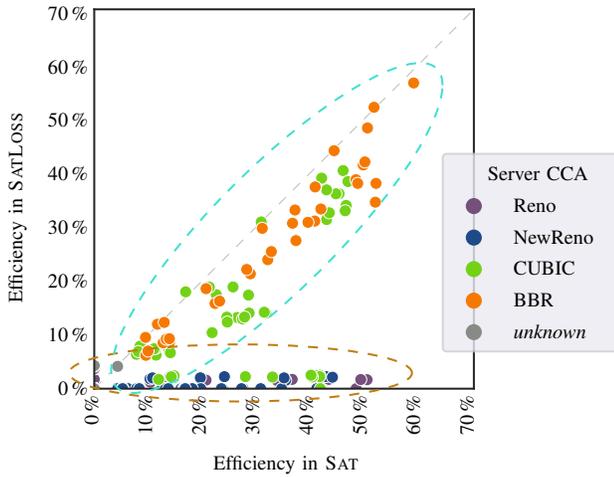}
    }
    \caption{%
        Relation of Measurement Results Between \sat/ \& \satl/\\
        Points are colored according to the \ac{CCA} of the server implementation.
    }%
    \label{fig:cca}
\end{figure}

As we assume that the \ac{CCA} of the server implementation has a large impact on the performance of a connection, we determined the supported algorithms and collected them in \cref{tab:impls_with_cca}.
For each implementation, we tried to identify all implemented algorithms and the default one by inspecting the available command line flags of the implementations in the \docker/ images used by the runner or the corresponding code if available.
As there is no database or standardized interface to query the \acp{CCA}, it is not guaranteed that the information in \cref{tab:impls_with_cca} is correct.
Additionally, it should be noted that the coding and parametrization of the same \acp{CCA} might differ substantially among implementations.

\Cref{fig:cca} displays a relational plot of the measurement results from the \sat/ scenario (horizontal) and the \satl/ scenario (vertical).
Again, we use the normalized efficiency values as defined in \cref{eqn:efficiency}.
Each point corresponds to the mean of the efficiencies achieved by the corresponding implementation tuple over all ten iterations in the corresponding scenario.
Two clusters of measurement results can be identified:
The first (\textcolor{Chocolate}{\faCircle[regular]} brown) one shows no correlation between both experiments.
While there are medium results in the \sat/ scenario, the results achieved in the \satl/ scenario (with a \ac{PLR} of \SI{1}{\percent}) are altogether very poor.
The other group (\textcolor{turquoise}{\faCircle[regular]} turquoise) shows a correlation between both experiments, and the results in \satl/ are quite similar and only slightly poorer than in \sat/.

We use the previously determined \acp{CCA} to color the points in that plot according to the algorithm used by the server implementation.
In the turquoise group of data points achieving a link utilization of more than \SI{10}{\percent} in the \satl/ scenario, there are only combinations where the server uses \CUBIC/ or \BBR/.
The other group contains combinations with \Reno/, \NewReno/ and \CUBIC/.

This could be an indication for \CUBIC/ and \BBR/ working better in lossy scenarios than \Reno/ and \NewReno/.
However, it might also be possible that implementations that employ one of the much more complex \acp{CCA} are more optimized in general.
It is hard to draw a more precise conclusion without deeper knowledge of the code of the implementations.

\subsection{Time-Offset Graphs}
\label{sec:detailed_graphs}

While we try to render and annotate time-offset plots using captured \pcap/ traces, this does not succeed for every combination for various reasons.
For example, some combinations do not use \UDP/ port 443, some use \HTTP//3 instead of \HTTP//0.9 as requested by the runner, or there are issues with standard compliance.
In this section, we chose five plots that show interesting behavior and discuss them in more detail.
More plots can be found on our website\textsuperscript{\ref{fn:qir_se_website}}.
In every plot there are the traces of all iterations of the same implementation tuple in the according measurement scenario (ten for \sat/ and \satl/, five for \astra/ and \eut/).
That one with the medium time to completion is highlighted colored.
The others are gray.
If an offset number is transmitted the first time, the corresponding point in the plot is colored \textcolor{SkyBlue}{\faCircle} blue.
Retransmissions are colored \textcolor{Orange}{\faCircle} orange.

\subsubsection{\picoquic/---\picoquic/---\satl/}
\label{sec:detailed_graphs_picoquic_picoquic_satl}

\begin{figure}
    \centering
    \includegraphics[width=\linewidth]{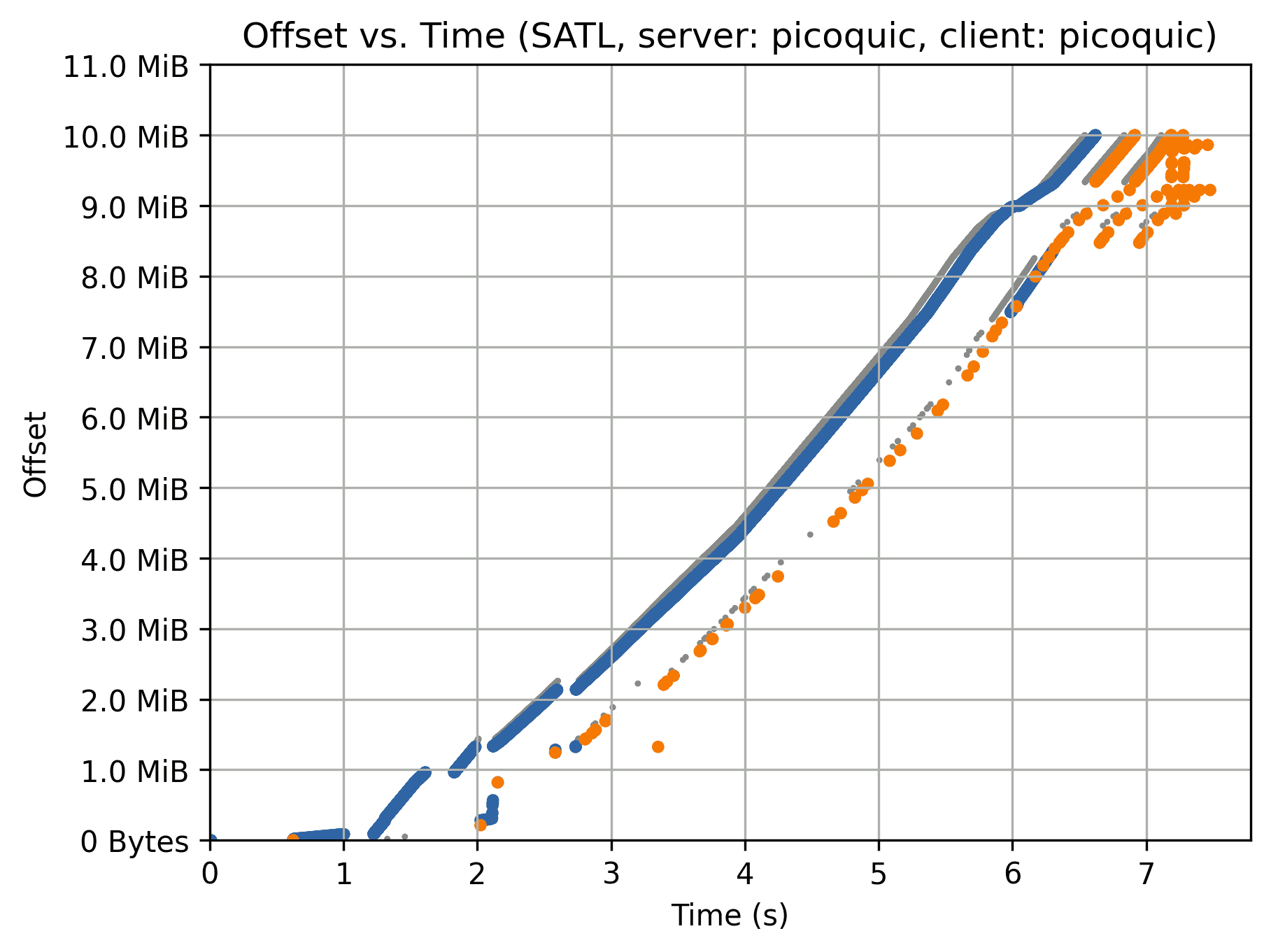}
    \caption{Time-Offset Plot of \Picoquic/---\Picoquic/---\satl/}%
    \label{fig:picoquic_trace}
    \vspace{1em}%
\end{figure}

Compared to the other implementations, \picoquic/ often achieves good results.
It seems to be built with satellite use-cases in mind.
Even in challenging scenarios, the transmission makes steady progress.
\Cref{fig:picoquic_trace} shows the transmission of \picoquic/ as client and as server in the \satl/ scenario---the simulated scenario with \SI{1}{\percent} \ac{PLR}.
According to our analysis in \cref{tab:impls_with_cca}, \BBR/ was used in this transmission.
As soon as the bottleneck bandwidth is estimated, the data is transmitted sequentially at a fairly constant rate and single lost packets are re-transmitted out-of-order. %
\Picoquic/ seems to automatically retransmit the last bytes of the file multiple times to avoid waiting one \RTT/ until the client can signal a loss.
The entire transmission takes slightly more than \SI{7}{\second}.
This equals a goodput of slightly more than \SI{11}{\mbps} which equals a link utilization of slightly more than \SI{50}{\percent}.

\subsubsection{\kwik/---\msquic/---\sat/}

\begin{figure}[t]
    \centering
    \includegraphics[width=\linewidth]{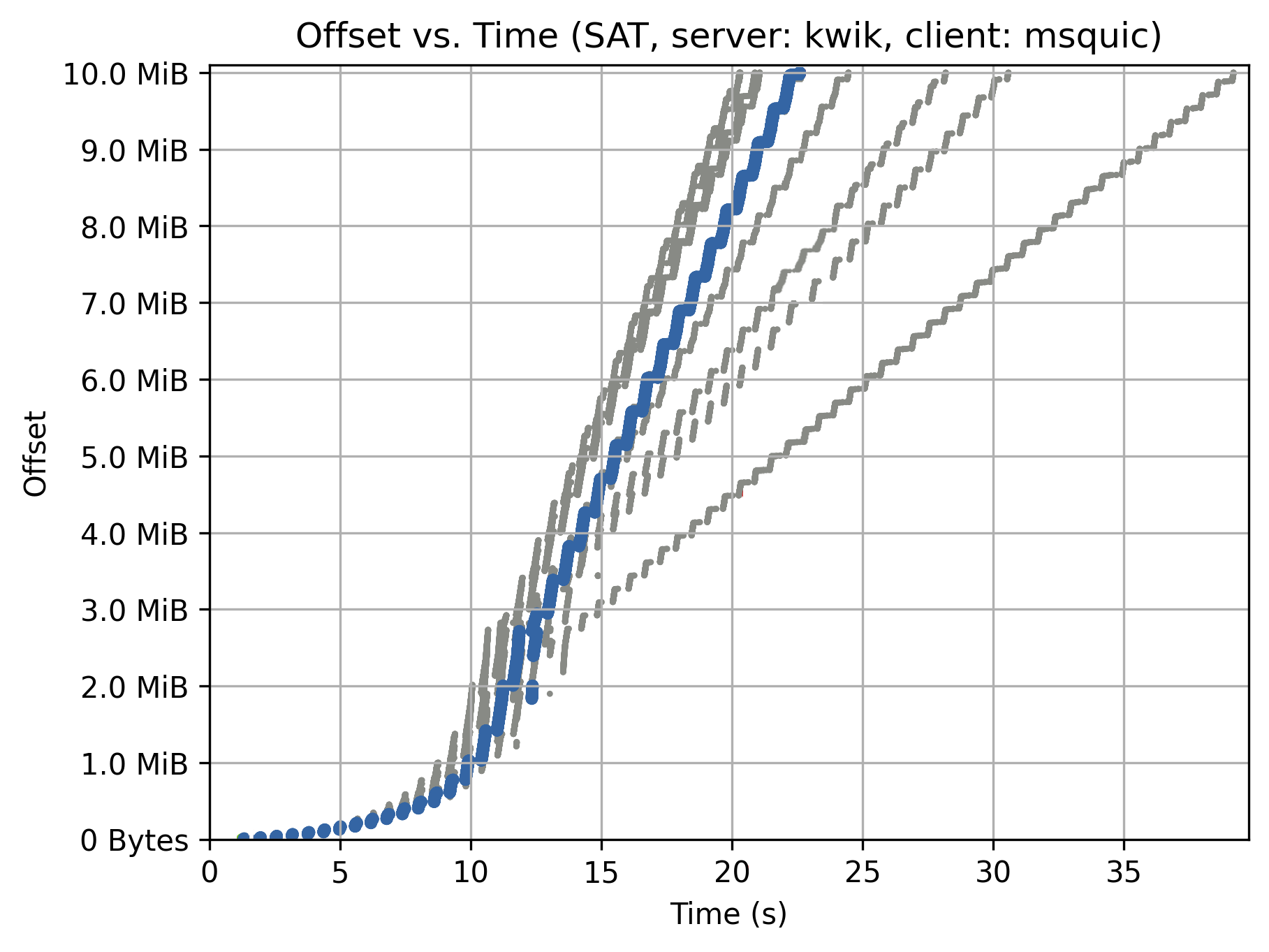}
    \caption{Time-Offset Plot of \Kwik/---\Msquic/---\sat/}%
    \label{fig:kwik_msquic_trace}
    \vspace{2.5em}%
\end{figure}

The example in \cref{fig:kwik_msquic_trace}, with \kwik/ as server using \NewReno/ as \ac{CCA}, shows a slow startup phase of about \SI{11}{\second}.
Most implementations that are not optimized for satellite scenarios have this in common.
Some implementations, like \aioquic/, even never reach a steady state.
In this example there is a second striking behavior:
In some iterations, the final transmission rate is set to a much smaller value than in other iterations.
This results in very different time to completions between iterations of the same experiment---even though no nondeterministic packet losses are emulated in the \sat/ test case.
Additionally, the jagged curves indicate bad pacing.

\subsubsection{\lsquic/---\xquic/---\sat/}

\begin{figure}[t]
    \centering
    \includegraphics[width=\linewidth]{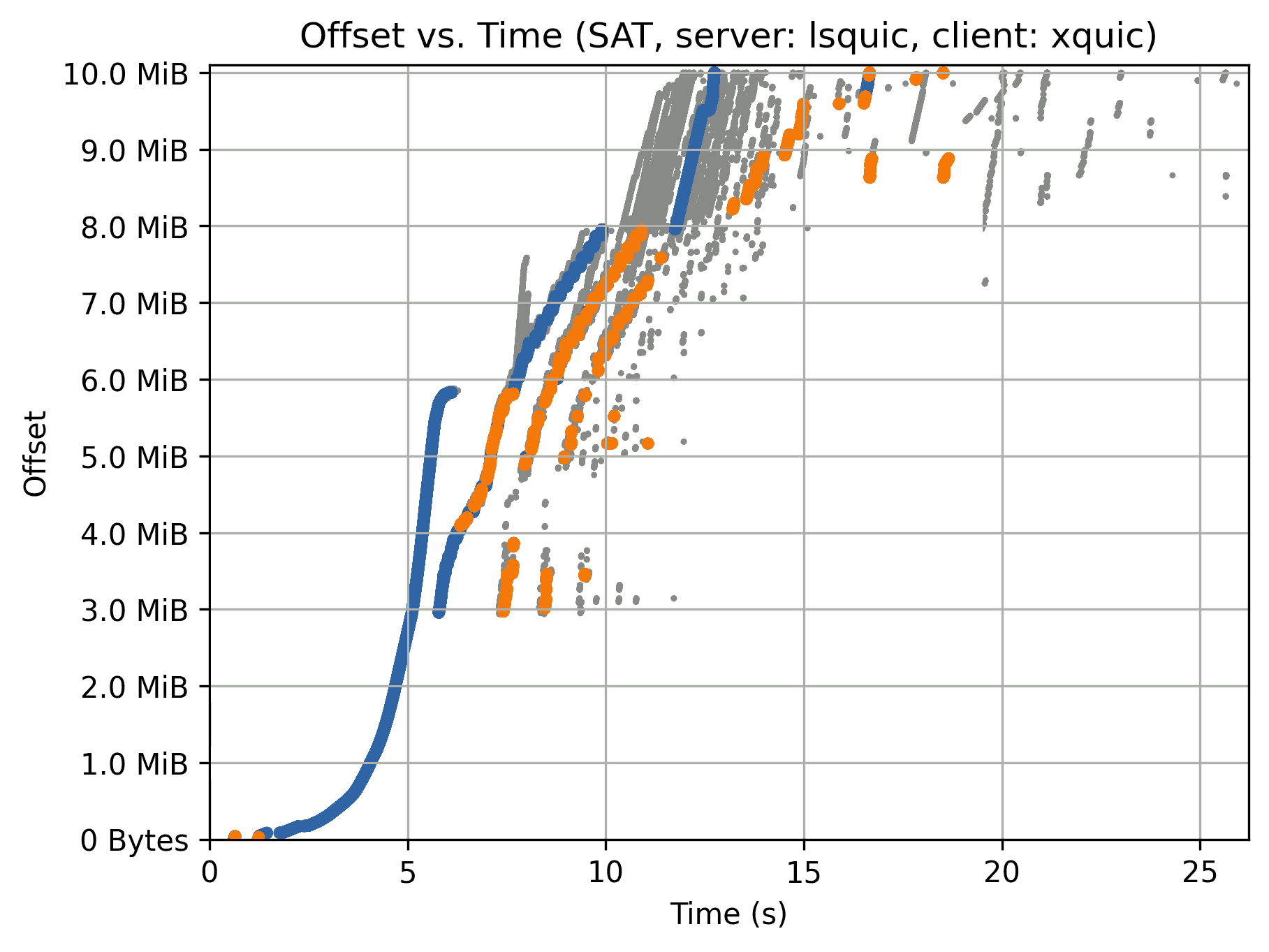}
    \caption{Time-Offset Plot of \Lsquic/---\Xquic/---\sat/}%
    \label{fig:lsquic_xquic_trace}
    \vspace{1em}%
\end{figure}

The analysis of the transmissions shown in \cref{fig:lsquic_xquic_trace} reveals remarkably many retransmissions.
Since no artificial losses are introduced in the \sat/ scenario, the retransmissions might indicate a bug in either of the implementations.
However, some sequence numbers in the range between 3 and \SI{4}{\mib} seem not to be used sequentially.
The second almost vertical line in this range is not drawn in orange, which indicates that the according packets are no retransmitted ones.
Such behavior should be analyzed in the future.

\subsubsection{\msquic/---\xquic/---\satl/}

\begin{figure}[t]
    \centering
    \includegraphics[width=\linewidth]{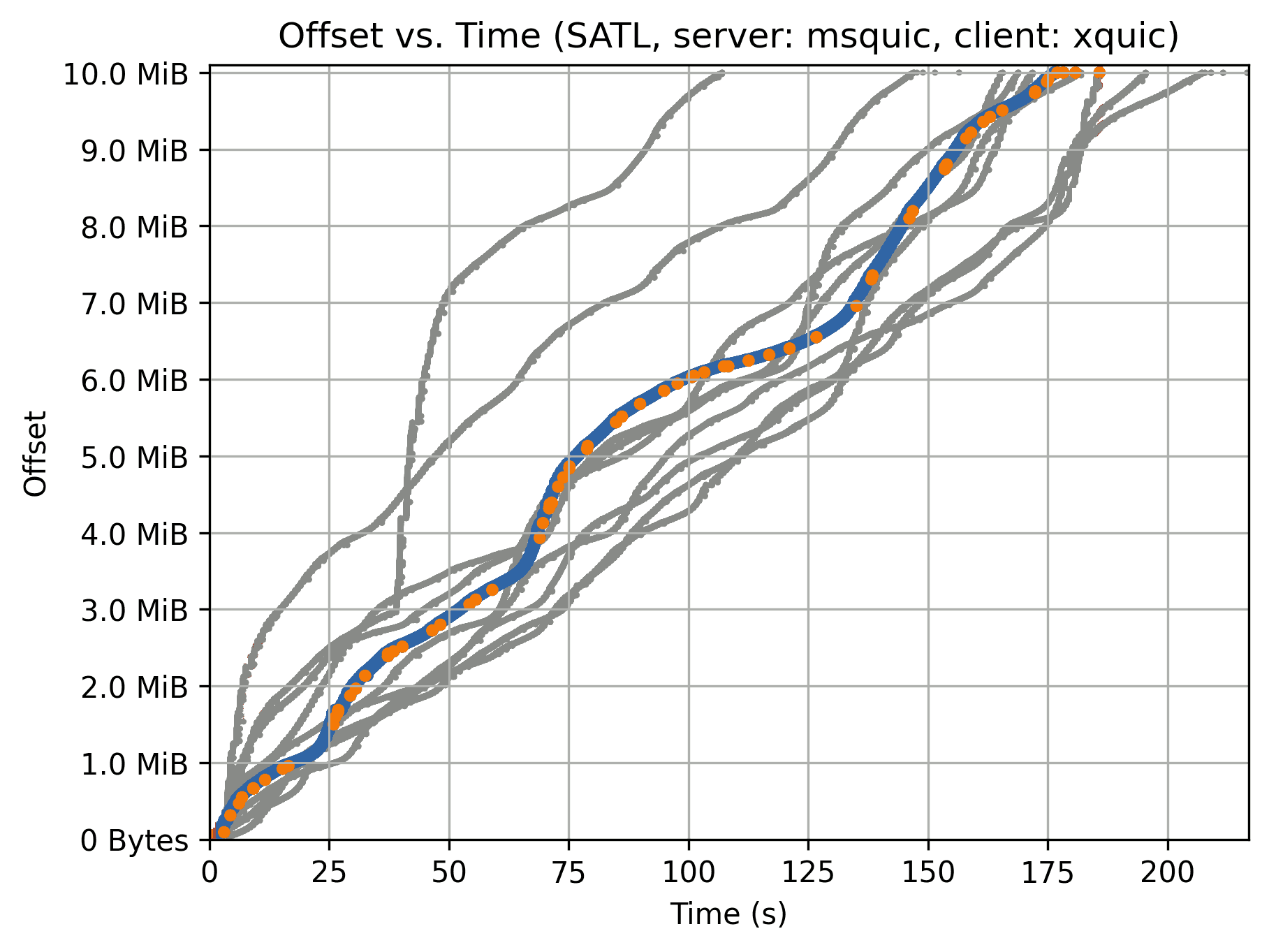}
    \caption{Time-Offset Plot of \Msquic/---\Xquic/---\satl/}%
    \label{fig:msquic_xquic_trace}
    \vspace{1em}%
\end{figure}

The plot in \cref{fig:msquic_xquic_trace} shows very different but high values for the time to completion in the \satl/ scenario.
According to \cref{tab:impls_with_cca}, loss-based \CUBIC/ was (most likely) used in this scenario.
The losses lead to high drops of the data rate and recovery seems to be very slow probably due to the high \RTT/.
When many losses occur in quick succession, the data rate is throttled more heavily.
When losses indicate congestion, it is fine to throttle the transmission rate.
However, when packets get lost due to temporary effects on a path segment, there is no benefit in reducing the transmission data rate.
Instead, doing so has a severe impact on the performance via the satellite link.

\subsubsection{\picoquic/---\ngtcp/---\astra/}

\begin{figure}[t]
    \centering
    \includegraphics[width=\linewidth]{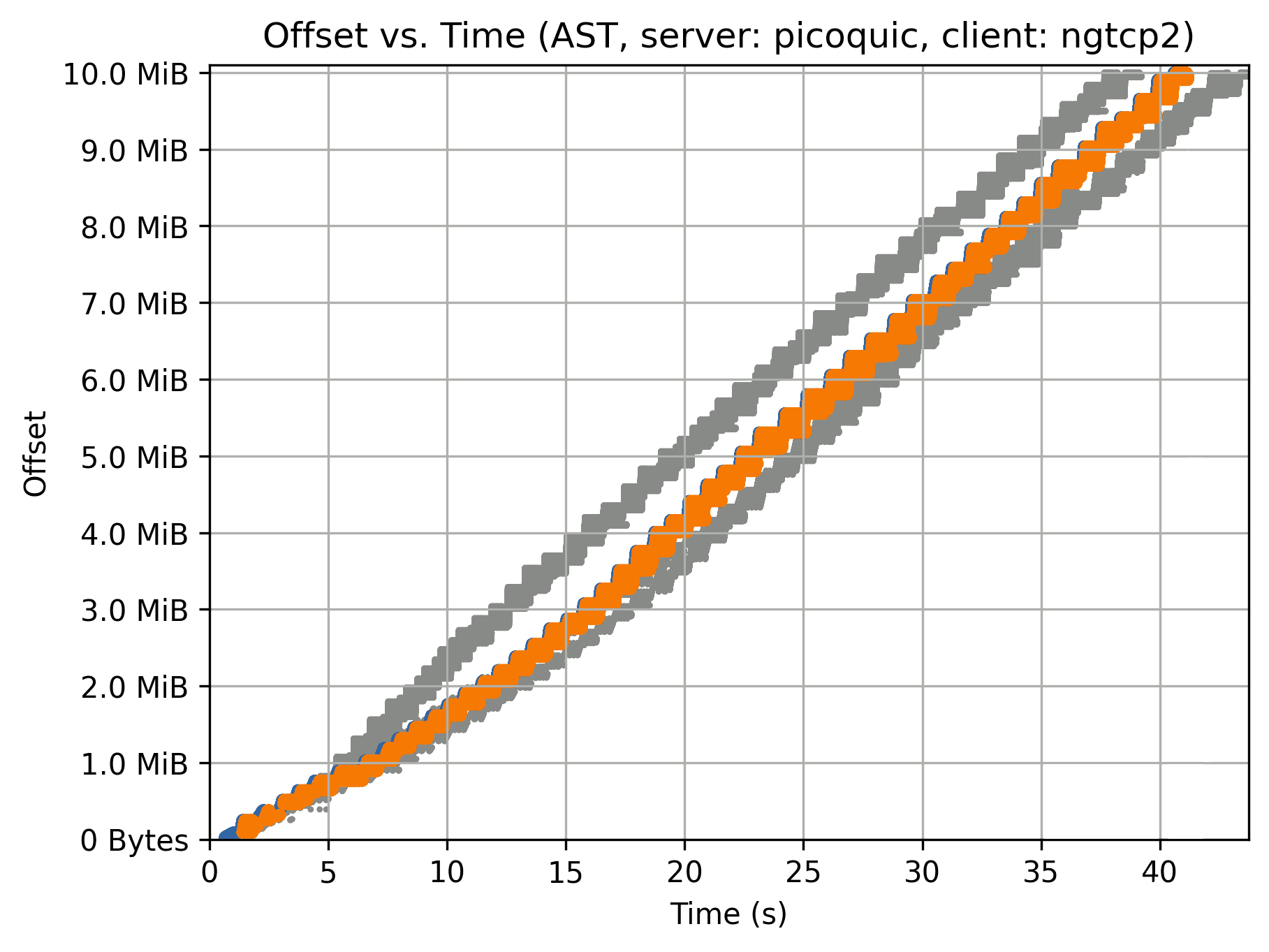}
    \caption{Time-Offset Plot of \Picoquic/---\Ngtcp/---\astra/}%
    \label{fig:picoquic_ngtcp2_ast}
\end{figure}

A way to avoid dreaded retransmissions is sending each packet multiple times even when it was not lost.
This can speed up the transmission because it is not necessary to wait for the \ACK/ of the recipient, which has proven to be beneficial in previous \cref{sec:detailed_graphs_picoquic_picoquic_satl} where \picoquic/ was used as server, too.
On the other hand, this optimization causes additional traffic on the link and thus should be used carefully.
In \cref{fig:picoquic_ngtcp2_ast}, it can be observed that such optimistic retransmissions leads to lots of redundantly sent data, because every packet is retransmitted six to eight times.
Although only a file of \SI{10}{\mib} is transferred, this leads to approximately \SIrange{60}{80}{\mib} of data being sent.

\pagebreak{}

\section{Conclusion and Future Work}%
\label{sec:conclusion}

Due to the non-applicability of \acp{PEP}, the performance of \QUIC/ via geostationary satellite links depends mainly on the end points.
We presented a tool based on the \ac{QIR} that allows running measurements via real and emulated satellite links with numerous \QUIC/ implementations.
It also automatically generates time-offset plots for more detailed analysis.
Our results show that differences between implementations are very high.
Many implementations even fail to transfer a medium-sized file over satellite links.
The successful ones usually achieve a very poor goodput.
The performance is even poorer when artificial packet losses are introduced.
Increasing the link data rate does not automatically increase the achieved goodput in the same ratio.
According to the results of the combinations of implementations, we can conclude that both the server and the client contribute to the overall performance of the transmission.

Regarding future work, multiple directions can be thought of.
Additional performance test cases (e.g., short transfers of small objects, enabling explicit congestion notification, fairness tests, etc.) would provide more insights into the performance of \QUIC/ over satellite networks.
This should also include different parameters for \QUIC/ protocol stacks~\cite{jonesEnhancingTransportProtocols2021} and extensions like the 0-RTT-BDP draft~\cite{kuhnTransportParameters0RTT2021}.
Testing could include more satellite systems, including both geostationary and \ac{LEO} megaconstellation satellite systems (e.g., Starlink).
Finally, long-term measurements would be helpful to track changes of \QUIC/ implementations.
This should also include investigations regarding the high number of unsuccessful runs observed in the \ac{QIR-SE}.

\section*{Acknowledgement}

We would like to thank \href{https://github.com/marten-seemann/quic-interop-runner/graphs/contributors}{Marten Seemann and all contributors} for developing and maintaining the \ac{QIR}.

\vspace{5ex}
\begin{minipage}{0.3\linewidth}
    \includegraphics[width=\linewidth]{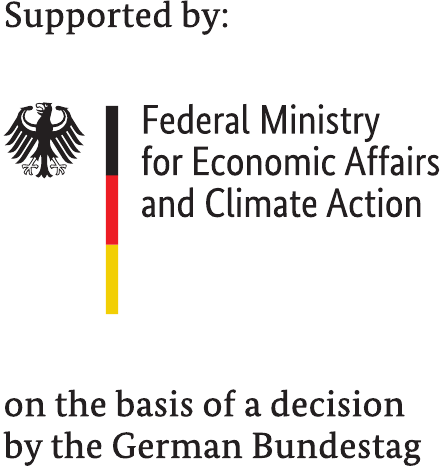}
\end{minipage} \hfill
\begin{minipage}{0.55\linewidth}
    This work has been funded by the Federal Ministry for Economic Affairs and Climate Action in the project QUICSAT.
\end{minipage}
\vspace{5ex}

\bibliographystyle{IEEEtran}
\bibliography{paper}

\mbox{%
    \centerline{\textit{All Internet links were last accessed on 2022-02-12.}}
}

\end{document}